\documentclass[letterpaper,10pt,journal]{IEEEtran}
\usepackage{graphicx} 
\usepackage{hyperref}
\usepackage{amsmath}
\usepackage{underscore} 
\usepackage{authblk}
\begin{document}

\title{Sheaf-Theoretic Causal Emergence for Resilience Analysis in Distributed Systems}
\author[1,2]{Anatoly A. Krasnovsky}
\affil[1]{Department of Computer Science and Engineering, Innopolis University}
\affil[2]{Q Deep, Innopolis, 420500, Russia}
\date{\today}
\maketitle

\begin{abstract}
Distributed systems often exhibit emergent behaviors that impact their resilience \cite{franz-kaiser-topological-2020,adilson-e-motter-cascadebased-2002,jianxi-gao-universal-2016}. This paper presents a theoretical framework combining \textit{attributed graph models}, \textit{flow-on-graph simulation}, and \textit{sheaf-theoretic causal emergence analysis} to evaluate system resilience. We model a distributed system as a graph with attributes (capturing component state and connections) and use sheaf theory to formalize how local interactions compose into global states. A flow simulation on this graph propagates functional loads and failures. To assess resilience, we apply the concept of \textit{causal emergence}, quantifying whether macro-level dynamics (coarse-grained groupings) exhibit stronger causal efficacy (via effective information) than micro-level dynamics. The novelty lies in uniting sheaf-based formalization with causal metrics to identify emergent resilient structures. We discuss limitless potential applications (illustrated by microservices, neural networks, and power grids) and outline future steps toward implementing this framework \cite{lake2015human}.
\end{abstract}

\section{Introduction}
Complex distributed systems---in computing, infrastructure, biological networks, and beyond---can exhibit resilience through emergent structural properties. Traditional resilience analyses often rely on graph-based metrics or cascading-failure simulations, which may overlook \emph{causal emergence}: situations where higher-level (macro) groupings of components yield more informative or deterministic dynamics than considering the components in isolation. Recent work in complex systems science has shown that macro-scale models can sometimes “beat” micro-scale models in terms of causal power \cite{erik-hoel-quantifying-2013,hoel-when-2017}, a phenomenon quantified using \emph{effective information} (EI) \cite{giulio-tononi-integrated-2016}.

In parallel, applied topology has demonstrated that \emph{sheaf theory} can model distributed systems by capturing local-to-global consistency constraints \cite{michael-robinson-sheaves-2016,robert-ghrist-elementary-2014,jakob-hansen-opinion-2021,gunnar-carlsson-topology-2009}. This paper merges these approaches. By combining sheaf-based representations with EI-based causal emergence metrics, we aim to identify when and how macro-scale structures in a distributed system contribute to overall resilience.

We present here the theoretical framework. We introduce (i) an \emph{attributed graph} model, (ii) a \emph{cellular sheaf} construction encoding node and edge data with consistency maps, (iii) a \emph{flow simulation} technique for studying failures, and (iv) an \emph{EI-based causal emergence} measure for evaluating resilience across different levels of system description. Our work is domain-agnostic, though we illustrate possible applications in microservices, neuroscience, and power networks.

\subsection{Related Work and Novel Contributions}

The use of sheaf-theoretic methods to study networks has been pioneered by Robinson, Ghrist, Hansen, and others \cite{michael-robinson-sheaves-2016, robert-ghrist-elementary-2014, jakob-hansen-opinion-2021}. In particular, these works leverage cellular sheaves to encode local-to-global constraints and to detect global consistency patterns in complex networks (see also \cite{Chazal-2017} for an introduction to topological data analysis). Related approaches in network science also address local data fusion or inference.

Our framework builds upon these foundations but offers two distinguishing features. First, we integrate EI–based causal emergence analysis with the sheaf structure. While previous efforts have largely focused on consistency of local information or topological invariants, we address \textit{how macroscopic causal patterns can arise} and relate directly to system resilience \cite{Seth-2010, Lizier-2012}. Second, our proposed \textit{flow-on-graph simulation} explicitly models node and edge failures, providing a concrete route to measure cause-and-effect relationships in distributed systems subject to perturbations.

This work therefore extends existing literature by uniting causal graphical methods with sheaf-based consistency, demonstrating how network-level resilience can be \textit{quantitatively} assessed through EI metrics. In doing so, we do not seek to “reinvent the wheel”, but rather to augment known sheaf-theoretic tools with a formally grounded measure of causal emergence.

\subsection{Problem Statement and Scope}

In this work, we address the following overarching question:

\textit{How can we 
formally detect and measure emergent resilience in a networked system, and under 
what conditions do macro-level descriptions provide stronger causal explanatory 
power than micro-level models?} 

Our theoretical framework systematically integrates: 
\begin{itemize}
    \item \textbf{Sheaf-based modeling} of local-to-global constraints:
    This apparatus encodes how local node/edge data must conform to consistency maps, thereby forming valid ``global sections.'' Through this approach, we capture the interplay between component 
    states and system-level behaviors;
    \item \textbf{Flow simulations} to track how failures or perturbations propagate;
    \item \textbf{Causal emergence metrics} (specifically, EI) 
          to compare the predictive and explanatory power of different levels of 
          system description\cite{Seth-2010,Lizier-2012}.
\end{itemize}

By quantifying causal emergence across these scales, we show how certain 
aggregations of nodes---whether microservices, neuronal assemblies, or power 
subnetworks---can exhibit higher resilience. Therefore, we provide a unified 
approach to understanding \textit{when and why} macro-level structures outperform 
micro-level descriptions in terms of robust functioning and causal efficacy.

\section{Theoretical Framework}

\subsection{System Modeling with Attributed Graphs}
A distributed system is represented by a directed graph $G=(V,E)$ with node set $V$ and edge set $E$. Each node $v\in V$ represents a component (a microservice, neuron, or grid bus), and each directed edge $e\in E$ a communication or functional link (API call, synaptic projection, transmission line). We associate with each node $v$ an attribute space $X_v$ (e.g., throughput, firing rate, voltage level) and with each edge $e$ an attribute space $X_e$ (e.g., latency, synaptic strength, electrical flow). A label function $L$ gathers these attributes into a structured data set for $v$ or $e$.

\subsection{Sheaf Construction for Local-Global Consistency}
We build a \emph{cellular sheaf} $\mathcal{F}$ over $G$. For each node $v$, $\mathcal{F}(v)$ is a data space of possible local states. For each directed edge $e:u\to v$, $\mathcal{F}(e)$ is a space capturing interaction data. Restriction maps $\rho_{e\to u}:\mathcal{F}(u)\to\mathcal{F}(e)$ and $\rho_{e\to v}:\mathcal{F}(v)\to\mathcal{F}(e)$ enforce consistency between endpoint states and the edge’s data, ensuring local compatibility across the network. A \textit{global section space}, denoted by $\Gamma(\mathcal{F})$, is the set of all assignments of data to the nodes and edges of $G$ that satisfy the restriction maps (i.e., all globally consistent configurations). Formally, 
\[
   \Gamma(\mathcal{F}) \;=\; 
     \{\, \sigma \mid \sigma \text{ respects each } \rho_{e \to u} 
       \text{ and } \rho_{e \to v}\}.
\]
A \emph{global section} $\sigma\in\Gamma(\mathcal{F})$ is an assignment of values to every node and edge that respects these maps, representing a globally consistent system state.

\subsubsection*{Comparison with Bayesian Networks}

It is important to note that \textit{Bayesian networks} and \textit{sheaf-theoretic models} interpret edges quite differently. In a Bayesian network, an edge from node $t$ to node $s$ indicates a conditional dependence of $s$ on a joint distribution over all its parent nodes \cite{koski2011bayesian}. Simultaneously, in a sheaf, each edge carries a \textit{separate restriction map}, specifying how the local state of one node restricts or aligns with another node’s data. Although these representations may seem superficially similar, they rely on different semantics. Bayesian edges encode \textit{probabilistic conditional dependencies}, whereas sheaf edges enforce \textit{consistency constraints} among local data spaces. When incorporating causal emergence metrics such as EI in a sheaf context, we must carefully keep these formalisms distinct, recognizing that joint parent-child mappings in Bayesian networks do not translate directly to individual restriction maps in a sheaf. Our proposed approach retains the sheaf-based view of local consistency while still quantifying mutual information and causal influence across nodes.

\subsection{Flow-on-Graph Simulation of Dynamics and Failures}
To rigorously capture how local changes propagate through the network, we introduce a flow-based formalism \emph{within the sheaf setting}. Here, each node and edge in the attributed graph not only stores static attributes but also participates in a dynamic process of transferring or transforming “flow” (e.g., data, energy, signals). By describing this flow in terms of sheaf-based consistency maps, we can track how local failures or adjustments force new global sections to emerge, thus revealing the mechanism by which resilient or cascading behaviors occur.

We model system behavior by simulating \emph{flows} over $G$ (e.g., data requests, power, neural signals). An initial global section $\sigma_0$ represents a normal operating point. We then introduce dynamics (discrete or continuous) to update node and edge states, ensuring at each step that they remain consistent with $\mathcal{F}$. Failures (e.g., node crashes, edge outages) are simulated by altering attributes in $X_v$ or $X_e$ (setting capacity to zero, forcing a node offline, etc.), after which the system finds a new configuration $\sigma_1$ that may reflect reduced functionality. Throughout, we record cause-effect relationships by systematically injecting controlled perturbations \cite{franz-kaiser-topological-2020,adilson-e-motter-cascadebased-2002,jianxi-gao-universal-2016}.

\subsection{Causal Emergence Analysis using Sheaves}
EI is a measure of how interventions on one set of variables affect another set of variables in a system. In the \textit{do-operator} framework of Pearlian causality \cite{pearl-causality-2009}, we define
\[
  \mathrm{EI} \;=\; I\bigl(X_{\mathrm{do}};\, X_{\mathrm{effect}}\bigr),
\]
where $X_{\mathrm{do}}$ represents the perturbed components and 
$X_{\mathrm{effect}}$ the resulting system states after those perturbations.

\subsubsection{Macro-Node Aggregation Criteria}
A distinguishing feature of our approach is identifying \emph{macro-nodes} via the sheaf. Specifically, we look for subgraphs $U\subseteq V$ whose internal nodes exhibit strong mutual dependencies or redundancy. One pragmatic selection rule is to search for $U$ with \emph{maximal pairwise EI} among its nodes, meaning the components in $U$ tend to synchronize or fail as a unit. Formally, we can define:
\[
  U^* = \arg\max_{U \subseteq V} \;\left(\sum_{u_i,u_j\in U}\mathrm{EI}(u_i;u_j)\right),
\]
where $\mathrm{EI}(u_i;u_j)$ measures the causal influence of node $u_i$ on $u_j$. Once identified, we \emph{collapse} $U$ into a single macro-node $v_U$ in an aggregated graph $G'$, preserving the edges to/from $U$ by mapping them to/from $v_U$. The sheaf is similarly collapsed via “quotient” or “sub-sheaf” constructions, resulting in a higher-level node whose internal states represent all consistent global sections over $U$.

\subsubsection{EI Computation}
To quantify causal influence at micro or macro levels, we adopt the “\textit{do}-operator” framework from Pearlian causality \cite{pearl-causality-2009}, measuring:
\[
  \text{EI} = I\bigl(X_{\mathrm{do}};\,X_{\mathrm{effect}}\bigr),
\]
where $X_{\mathrm{do}}$ is the system variable(s) we perturb, and $X_{\mathrm{effect}}$ is the resulting system state after the perturbation. In practice:

\begin{itemize}
    \item We enumerate (or sample) possible interventions on $X_{\mathrm{do}}$.  
    \item For each intervention, we run the flow simulation to obtain the outcome distribution of $X_{\mathrm{effect}}$.  
    \item We estimate the mutual information $I(X_{\mathrm{do}}; X_{\mathrm{effect}})$ from these outcomes.  
\end{itemize}

\textbf{Algorithmic Outline:}
\begin{enumerate}
    \item \textbf{Input:} Graph $G=(V,E)$, sheaf $\mathcal{F}$, initial global section $\sigma_0$.
    \item \textbf{Select Variables to Perturb:} Identify $X_{\mathrm{do}}$ (e.g., node states).
    \item \textbf{Enumerate Interventions:} For each candidate state/configuration $d_i\in X_{\mathrm{do}}$, apply $d_i$ to the system (simulate a “do($d_i$)” operation).
    \item \textbf{Simulate Outcomes:} Evolve the system under each intervention to yield $x_{\mathrm{effect}}(d_i)$.
    \item \textbf{Estimate EI:} Compute $I(X_{\mathrm{do}}; X_{\mathrm{effect}})$ across all interventions.
    \item \textbf{Compare Micro vs.\ Macro:} Repeat at the aggregated (macro-node) level; measure any increase in EI to detect causal emergence.
\end{enumerate}

If the macro-level EI exceeds the micro-level EI, we say the system exhibits \emph{causal emergence} at that grouping, suggesting a resilient macro-scale structure. We define a \emph{causal resilience index} $R_{\mathrm{cause}}=\mathrm{EI}_{\text{macro}}-\mathrm{EI}_{\text{micro}}$. A positive $R_{\mathrm{cause}}$ implies that the aggregated subgraph is more informative about future states than the individual nodes in isolation \cite{erik-hoel-quantifying-2013,hoel-when-2017}.

\begin{figure}[!t]
\centering
\includegraphics[width=\columnwidth]{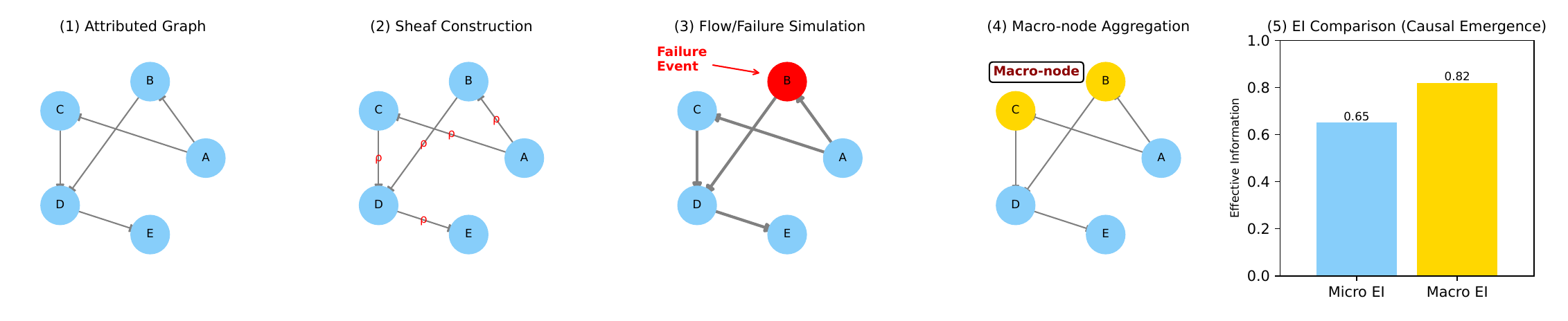}
\caption{Schematic Overview of the Proposed Framework. 
(1) Represent the system as an attributed graph. 
(2) Construct a cellular sheaf for local-to-global consistency. 
(3) Simulate flows and failures. 
(4) Aggregate subgraphs into macro-nodes. 
(5) Compare micro-level and macro-level EI to detect causal emergence.}
\label{fig:workflow}
\end{figure}

\section{Potential Applications}
Although this framework is generally applicable to any distributed system, we illustrate with three common domains.

\subsection{Cloud Microservices}
Each microservice is modeled as a node, with edges for communication links. The sheaf enforces consistency across service interfaces. Flow simulation captures how requests propagate under normal and failed conditions. Causal emergence occurs if a group of services consistently “succeed or fail” together, forming a resilience boundary. This can inform architectural decisions about fault tolerance or scaling strategies.

\subsection{Computational Neuroscience}
Neurons or brain regions form nodes, edges track synapses or fiber tracts, and the sheaf encodes constraints (e.g., conservation of signals). Lesion simulations reveal how activity reroutes. A cluster of neurons might exhibit stronger macro-level causal power than the sum of single neurons, indicating a robust “assembly” that persists despite partial failure \cite{leon-o-chua-cellular-1988,giulio-tononi-integrated-2016,sizemore-importance-2019}.

\subsection{Power Grid Networks}
Nodes represent generators/loads/substations, edges are transmission lines, and the sheaf enforces flow or Kirchhoff constraints. Outage simulations identify stable islanding or rerouting behaviors. If a subset of buses can reliably maintain operation when partially disconnected, it may show higher EI as a macro-node, aligning with known microgrid resilience strategies \cite{franz-kaiser-topological-2020,adilson-e-motter-cascadebased-2002,jianxi-gao-universal-2016}.

\section{Future Work}
\subsection{Empirical Validation Timeline}
We plan to implement a toy microservice model in a simulation environment (e.g., \texttt{Python}/\texttt{NetworkX} plus a sheaf-based library \cite{michael-robinson-sheaves-2016}) to test causal emergence. We will compute $R_{\mathrm{cause}}$ under various failure scenarios and compare it to conventional resilience metrics (e.g., service response times, number of nodes surviving). If results correlate, it validates that higher EI indeed signals resilience in real architectures.

\subsection{Scalability Solutions}
A major challenge is the exponential number of possible interventions in large systems. We will explore:
\begin{itemize}
    \item \textbf{Approximate Sheaf Constructions:} Many real systems have repetitive or modular constraints that can be compressed in the sheaf representation.
    \item \textbf{Graph Sampling/Clustering:} Instead of simulating every node, we can sample representative subsets or apply hierarchical clustering to reduce complexity. 
    \item \textbf{Neural Approximations:} Building on \cite{cristian-bodnar-neural-2022}, we may use graph neural networks that incorporate sheaf constraints to learn approximate EI values.
\end{itemize}

\subsection{Advanced Theoretical Directions}
We intend to formalize quotient sheaves for macro-node aggregation, prove relationships between $R_{\mathrm{cause}}$ and topological invariants (e.g., sheaf cohomology \cite{michael-robinson-sheaves-2016,robert-ghrist-elementary-2014}), and extend the approach to time-evolving or asynchronous systems. Domain-specific adaptations (e.g., incorporating power flow solvers for power grids, spike-train analysis for neurons) will also be pursued.

\subsection{Future Directions and Universality of the Framework}
Although the preceding sections focused on resilience as a compelling application, 
the proposed combination of sheaf theory and causal emergence is more broadly 
applicable to establishing multi-level causality in distributed systems. By leveraging 
sheaf-based consistency maps and flow simulations, we can systematically capture 
local-to-global dependencies in a variety of domains—ranging from microservices 
architectures and neuroscience to power grid operation. 

In addition to evaluating robustness and resilience, this framework lays a foundation 
for discovering genuinely emergent causal structures. For instance, EI can be used to identify where coarse-grained groupings of nodes 
yield stronger causal explanations than a purely microscopic view \cite{Seth-2010, Lizier-2012}. Rigorous formal 
analysis, drawing on the algebraic properties of sheaf cohomology and on established 
causal inference principles, opens the door to theorem-driven insights about when 
and why macro-level causality might dominate in certain system configurations \cite{Shanmugam-2018}. Future 
work thus includes both rigorous theoretical generalizations (e.g., proving sufficient 
conditions for the emergence of higher EI at macro scales) and empirical studies to 
test these phenomena in practice. Through these endeavors, we hope to demonstrate 
that the proposed framework is not merely a tool for enhancing resilience analysis, 
but a \textit{general methodology} for examining how causal processes unfold across multiple 
layers of complex, networked systems.

\section{Conclusion}
We presented a theoretical framework that fuses \emph{sheaf theory} and \emph{causal emergence analysis} to investigate resilience in distributed systems. By representing a system as an \emph{attributed graph} and constructing a corresponding \emph{cellular sheaf}, one can rigorously track local-to-global constraints. Flow simulations then reveal how failures propagate, while \emph{EI} metrics quantify causal impact at multiple scales. A higher macro-level EI compared to micro-level EI signals \emph{causal emergence}, indicating resilient substructures. Although illustrated with microservices, neuroscience, and power grids, the methodology has limitless applicability to diverse networked systems. Future work will solidify this approach through empirical prototypes, scalability strategies, and deeper theoretical grounding, with the potential to transform our understanding and design of resilient distributed architectures.

\section*{Acknowledgment}
The author gratefully acknowledges Anton Ayzenberg and German Magai from Noeon Research for their constructive criticism and valuable advice, which guided the improvements to this work.

\bibliographystyle{IEEEtran}
\bibliography{main}

\end{document}